\documentclass{elsart}
\usepackage{amsmath,amssymb}
\usepackage{graphicx}
\usepackage{bm}

\begin{document}

\begin{frontmatter}
  \title{Rectification of motion in nonlinear media with asymmetric random drive} 

  \author{G.  Cebiroglu~$^{1,2}$, C. Weber~$^{1}$, and
    L.~Schimansky-Geier~$^{1}$}

  \address{$^1$~Institut f\"ur Physik, Humboldt--Universit\"at zu
    Berlin, Newtonstr. 15, 12489 Berlin, Germany, E-mail:
    alsg@physik.hu-berlin.de}

  \address{${^2}$~Institut f\"{u}r Mathematik, Humboldt-Universit\"{a}t
  Berlin, Rudower Chaussee 25, 12489 Berlin, Germany}

\begin{abstract}
  We consider moving particles in media with nonlinear friction and
  drive them by an asymmetric dichotomic Markov process.  Due to
  different energy dissipations, during the forward and backward stroke,
  we obtain a mean non-vanishing directed flow of the particles.
  Starting with the stationary velocity distribution, we calculate the
  stationary current of particles, the variance and the relative variance
  in dependence on the degree of nonlinearity of the friction, on
  the asymmetry and for different strengths of friction. In two
  dimensions the particle performs diffusional motion, if in addition
  the direction of the asymmetric drive changes stochastically.
\end{abstract}

\end{frontmatter}

 

\noindent
Devoted to P. H\"anggi on occasion of his 60th birthday.

\section{Introduction}
The occurrence of directed flows of matter or single particles, in
case of mean vanishing forces, is always connected with a break of
symmetry in the considered medium or device. Since the pioneering work
of H.  Purcell \cite {Purcell} a lot of work has been devoted to this
problem. Purcell found that simple devices require more than one
degree of freedom by which they can temporally store energy
sequentially in an asymmetric way. Later on, theoretical work on
ratchets has elaborated the many different combinations of possible
constructions between nonlinear periodic potentials, noise and forcing
to obtain non-vanishing flows of particles if the averaged sum of
forcing vanishes
\cite{magnasco,hangrev,astumian,PReiPHan02,Linke,anishchenko}. It was
P.  H\"anggi who put forward the notion of {\it Brownian motors}
\cite{hangrev,brownmotors,physicstoday} to point out the possibility
to find engines in the presence of a strongly fluctuating environment.
Most strikingly, in many cases this noisy environment can even act as
the source of energy which drives the engine (the interested reader
is referred to \cite{hanggi} for Peter H\"anggi's extended list of
publications on Brownian motors and the various different topics he
has investigated).

In this paper we study particles moving in a medium with
nonlinear friction \cite{denisov02,LsgEbErd05,Lin08,JSEbEwaLsg09}.
Therefore, the rate of dissipation will depend on their
velocity.  Achieving different values of velocity will
result in a different speed of dissipation. We will drive such
particles by random forces with a vanishing average.  We assume a two
state random telegraph signal or a dichotomous Markov process as force
acting on the particle \cite{gard,HorLev84,bena,dybiec}.  Crucially, we will
assume that this force acts asymmetrically, i.e. its two values in the
two directions will be different. Such an applied asymmetric force
elongates the velocity of the particle differently during the forward
and backward stroke.  Hence, it will dissipate the
transmitted energy during both strikes differently. For this reason the particle
is able to perform a directed motion even in the case if the applied
drive vanishes on average.

We will assume that the random drive is a time homogeneous stochastic
process. Hence, the problem becomes stationary in the long time limit
and we can use tools from the theory of stationary stochastic
processes. As a result we calculate the mean velocities and the
variance of the velocities for the one dimensional case. In two
dimensions, we add a stochastic rotation of the direction
of the applied asymmetric forcing and show first numeric results for
the diffusional motion of the particle.

We mention that similar problems have been discussed within the
framework of vibrational dynamics by Blekhman (see e.g.
\cite{Blekh}), who has used an alternative approach in the case of a
fast periodic driving to find asymptotic velocities of gliding
particles in many different situations. This approach is based on the
separation of time scales and averaging over the fast time scale of
the driving force. Similar techniques have also been used as
approximate solutions to elaborate instabilities in driven
oscillators, see e.g.  the Kapiza-(Magnus)pendulum in \cite{Landau}.
An application to our stochastic driving is also possible, but here we
rely on techniques using the stationary probability density for the
particle's velocity.

\section{Nonlinear friction and asymmetric forcing}
We study singular particles with unit mass $m=1$ which move in a
fluid. Therein the particles experience a frictional force $f(v)$
which damps the motion. To compensate the damping, the particles are
forced by additive temporal forces $\xi(t)$. We assume that this force
possesses a vanishing mean and stationary correlations
\begin{equation}
  \label{eq:force3}
  \left\langle \xi(t)\right\rangle = 0 \,,~~~  \left\langle \xi(t)\xi(t^\prime)\right\rangle = K(t-t^\prime) \quad.
\end{equation}

Thus the velocity $v(t)$ of the particles obeys the following
dynamical equation
\begin{equation}
  \dot{v}\left(t\right)=f\left(v(t)\right) +\xi(t)\quad. 
\label{mod1}
\end{equation}
The time-dependent force $\xi(t)$ drives the system permanently out of
(thermal) equilibrium. The velocity distribution is non-Maxwellian.
We underline that there is no spatial dependence in the description
(\ref{mod1}). This typical ingredient of Brownian ratchets, namely
spatially periodic gradients, is absent in our model. Thus the
rectification process of (\ref{mod1}) differs from the usual Brownian
ratchet mechanism \cite{hangrev}. As outlined, symmetry is broken due
to the asymmetric forcing realizing an asymmetric energy dissipation by
the nonlinear friction.

As driving force we will consider a random telegraph signal $\xi(t)$
which is a time-homogeneous Markov process. It can take two values
$\left\lbrace A_+,A_-\right\rbrace $ with constant transition rates
$\kappa_-/\kappa_+$ between the two states. $\kappa_+$ denotes the
rate of passage from $A_+$ to $A_-$ and analogously, $\kappa_-$
denotes the rate of passage from $A_-$ to $A_+$. The resulting time correlation function reads
\begin{equation}
 \langle \xi(t) \xi(s)\rangle=\frac{(A_+-A_-)^2\kappa_+\kappa_-}{(\kappa_++\kappa_-)^2}\cdot e^{-(\kappa_++\kappa_-)|t-s|}\quad .
\end{equation}

Particular interest
will be paid to unbiased dichotomous Markov processes, i.e.  processes
with a vanishing mean $\left\langle \xi(t)\right\rangle =0$, which
leads to
\begin{equation}
 A_+\kappa_-+A_-\kappa_+=0\quad.
\label{dmn12}
\end{equation}

The condition (\ref{dmn12}) reduces the amount of independent
parameters to three.  We therefore introduce the following set of
three independent parameters of the unbiased DMN process
\begin{align}
  &A=|A_-|+|A_+|,~~~\tau=\tau_-+\tau_+,\nonumber\\
  &0<p=\frac{\min(|A_-|,|A_+|)}{\max(|A_-|,|A_+|)}<1 ,
\label{dmn20}
\end{align}
where the mean waiting times $\tau_{\pm}$ for the states $A_{\pm}$
relate to the rates as $\tau_{\pm}=1/k_{\pm}$. The parameter $A$
measures the strength of the DMN process. $\tau$, the
mean time of one cycle, is the characteristic time scale of the DMN. It differs from the correlation
time $\tau_C=(\kappa_++\kappa_-)^{-1}$. The parameter $p$ will be the desired
parameter controlling the asymmetry of the driving. Symmetric driving
is located exactly at $p=1$. High asymmetries, i.e. when the stroke
amplitudes $|A_-|$ and $|A_+|$ differ much, let $p$ tend to zero.

We can express the variance of the DMN in terms of the new parameters,
resulting in
\begin{equation}
  var\left\lbrace \xi \right\rbrace=\langle \xi^2(t) \rangle =\frac{A^2 p}{(1+p)^2}\quad .
\end{equation}
The variance increases with the driving amplitude $A$. This is
not rather surprising. However, while looking at the variance profile for
a varying asymmetry $p$, one observes, that the variance reaches its
maximum when the driving is totally symmetric. On the other hand,
growing asymmetry (i.e. lower values of $p$) reduce the variance. In
the asymmetric limit $p\to 0$, the variance vanishes completely.

In particular we are interested in the effect of nonlinear friction.
Therefore it will be convenient to introduce a model of friction that
can model nonlinearity via a parameter, in order to easily control the
impact of nonlinearity on the system (\ref{mod1}). Prominent types of
friction, are the so called Stokes friction ($f(v)=-\gamma_s v$),
which is typical for small particles moving at low velocities through
viscous fluids, and the ``quadratic drag force'' ($f(v)=-\gamma_q
|v|v$), corresponding to situations of objects moving at relatively
large velocities through fluids, a widely occurring situation in
aerodynamical engineering. Another prominent type of friction is the
so called Coulomb friction ($f(v)=-\gamma_c\frac{v}{|v|}$). This form
of friction characterizes sliding frictions, the form of friction that
occurs, when two surfaces slide against each other. This situation is
sometimes called ``dry friction''.  The respective friction
coefficients $\gamma_i$ depend specifically on the characteristics
of the fluid and of the corresponding particles.

Moreover, Stokes friction is a linear friction model, while the other
two models, quadratic and Coulomb friction, are nonlinear.
Our model of friction $f(v)$ should be able to model more general
forms of friction, but at the same time able to reproduce the mentioned
three prominent models of friction as specific limit cases. An immediate candidate for such a
model is apparently
\begin{align}
f(v)&=-\gamma |v|^n \mbox{sign}(v)& \gamma>0,\quad n\in\mathbb{R}^+\quad .
\label{fri1}
\end{align}
This friction model introduces two parameters, namely the friction
coefficient $\gamma$ and the friction exponent $n$, allowing to model
a wide range of relevant types of friction beyond the introduced
cases ($n=0,1,2$). The friction of general fluids, that are
omnipresent in small scaled biological systems, may therefore be properly
modeled by (\ref{fri1}) (see e.g. \cite{Blekh}). 

In the other limit, for $n \to \infty$, (\ref{fri1}) reduces to
\begin{equation}
  f_{\infty}(v)=\lim_{n\to\infty}f(v)=\begin{cases}
    +\infty& \text{for $v>1$},\\
    0& \text{for $\|v\|<1$},\\
    -\infty& \text{for $v<-1$} .
\end{cases}\label{nun}
\end{equation}
The motion of the particle is constraint to the velocities between
$-1$ and $1$, since the damping friction forces (\ref{nun}) approaches
infinity for $|v|>1$. In the region $|v|<1$ the particle moves freely,
without damping, while at the height of the barrier sites $|v|=1$, the
particle gets reflected back to lower velocities. This is in some
sense similar to a particle moving in a potential well, with infinite
slopes of the respective walls. Note that in this case there is no
dependence on any friction constant $\gamma$ anymore.

\section{Stationary currents: Analytically tractable models}
In general, it is impossible to obtain an exact solution of, for
instance, the stationary density of a system, if an arbitrary form of
noise, especially colored noise, is considered. However, the simple
structure of the DMN, as a Markovian two-state process, enables us to
obtain exact and explicit results in terms of stationary distributions
\cite{HorLev84,dybiec,KiHorLefIn,KiHorLef}. The stationary solution $P_{st}$ for Eq.(\ref{mod1}) reads
\begin{align}
  P_{st}\left(v\right)&={N  |i_+(v)i_-(v)|}\\
  &\times\exp\biggl[-\kappa_+\int^v i_+(s)ds-\kappa_-\int^v i_-(s)ds\biggr],\nonumber
\label{pdis22}
\end{align}
where the functions $i_{\pm}(v)$ are defined as
\begin{equation}
 i_{\pm}(v)=\frac{1}{-\gamma|v|^n \textmd{sign}(v)+A_\pm} \quad .
\label{supp1}
\end{equation}

As one can see with Eq.(\ref{mod1}) and (\ref{fri1}), the
support of the stationary distribution is a compact interval
$[v_-,v_+]$, with $v_+=\sqrt[n]{A_+/\gamma}$ and $v_-=-\sqrt[n]{A_-/\gamma}$,
and beyond this interval the
stationary distribution has to vanishes.

In order to compute the integrals in Eq.(9), we look at the
general expression $\textstyle \frac{1}{a x^n+C}$ , for arbitrary
$a,C,x,n\in\mathbb{R}$. Notice that we can always write
\begin{equation}
\begin{split}
\frac{1}{a x^n+C}=\frac{1}{C}\sum_{i=0}^{\infty}(\frac{a x^n}{C})^i,
\label{supp3}
\end{split}
\end{equation}
as far as $\tfrac{a x^n}{C}<1$ holds. 

Integrals of this expression can be expressed by the the standard
classical (or Gaussian) hypergeometric series $_2\mathbb{F}_1$. In
its general form, it is defined as
\begin{equation}
_2\mathbb{F}_1\left[a,b,c,z\right]=\sum_{n=0}^{\infty}\frac{(a)_n(b)_n}{(c)_n}\frac{z^n}{n!},\qquad \mbox{for $a,b,c,z\in\mathbb{C}$}\quad .
\end{equation}
Integration of Eq.(11) yields
\begin{equation}
\begin{split}
\int^z \frac{dx}{a x^n+C}&=\frac{x}{C}\,_2\mathbb{F}_1\left[\frac{1}{n},1,1+\frac{1}{n},\frac{a}{C}z^n\right]
\end{split}
\label{prelim30}
\end{equation}
and consequently 
\begin{equation}
\begin{split}
  \int^v i_\pm(v)dv=\begin{cases}&\displaystyle{\frac{v}{A_\pm}}\,_2\mathbb{F}_1\left[\frac{1}{n},1,1+\frac{1}{n},-\frac{\gamma v^n}{A_\pm}\right]\,~~ v>0,\\\\
    &\displaystyle{\frac{v}{A_\pm}}\,_2\mathbb{F}_1\left[\frac{1}{n},1,1+\frac{1}{n},\frac{\gamma(-v)^n}{A_\pm}\right]\,v<0.\nonumber\\
\end{cases}
\end{split}
\end{equation}

Stationary moments can be computed straightforwardly. However, in general
it will be difficult to give an analytic closed formula for the
stationary moments. That is why we will first look at special cases and
discuss the general case of arbitrary $n\in\mathbb{R}^+$ in the context of simulations later on.

Let us start with the dry friction limit ($n=0$), i.e.
\begin{equation}
\dot{v}=-\gamma \ \mbox{sign}(v)+\xi(t)\quad .
\end{equation}
Notice that for the case $\gamma>max(|A_-|,|A_+|)$ no motion can occur
at all, since the critical force in order to induce motion is
$\pm\gamma$. In this
case the stationary distribution $P_{st}^0$ takes the form
\begin{equation}
P_{st}^0(v)=\delta(v), 
\label{pathodry}
\end{equation}
where the $\delta$ function denotes the Dirac distribution. 

To exclude this situation, we require
$\gamma<min(|A_-|,|A_+|)$. Then the critical friction force is
overcome in both situations, in the state $A_+$ as well as in the state $A_-$
of the DMN $\xi(t)$.

The functions $i_\pm(v)$ simply become constants which results in
\begin{equation}
 \begin{split}
   P_{st}^0(v)&=-N\begin{cases}
     \dfrac{\exp\biggl[-\dfrac{\kappa_+ v}{-\gamma+A_+}-\dfrac{\kappa_- v}{-\gamma+A_-}\biggr]}{\left(-\gamma+A_+\right)\left(-\gamma+A_-\right)}&\, v>0,\\\\
     \dfrac{\exp\biggl[-\dfrac{\kappa_+ v}{\gamma+A_+}-\dfrac{\kappa_-
         v}{\gamma+A_-}\biggr]}{\left(\gamma+A_+\right)\left(\gamma+A_-\right)}&\,v<0.
\end{cases}
\end{split}
 \label{pdis5}
\end{equation}
The stationary distribution $P_{st}^0$ is an exponential function for
every half line ($v>0$, $v<0$). Note that for (\ref{pdis5}) to be
normalizable, the exponentials should converge to zero
in the limits of infinity $\textstyle v\to\pm\infty$.  Therefore it
is sufficient to show, that the exponentials on every half line are
decaying, which is in accordance with our assumptions $A_+>0$,
$A_-<0$, $\kappa_\pm,\gamma>0$ and $\gamma<\min(|A_-|,|A_+|)$.

If the driving is asymmetric, we get two different 
limits for the distribution function at
the origin $v=0$, since
\begin{equation}
\lim_{v\to0^+}P_{st}^0(v)=\frac{N}{\left(-\gamma+A_+\right)\left(-\gamma+A_-\right)}
\end{equation}
and the left side limit yields
\begin{equation}
  \lim_{v\to0^-}P_{st}^0(v)=\frac{N}{\left(\gamma+A_+\right)\left(\gamma+A_-\right)}\quad.
\end{equation}
Thus, the stationary distribution is
discontinuous at $v=0$ in the dry friction limit ($n=0$). 

For the normalization constant $N$ we get
\begin{equation}
 N=-\frac{\gamma \kappa}{2}\quad .
\end{equation}
with $\kappa=\kappa_+ + \kappa_-$. The first moment, as the mean average velocity and stationary current
$J=\textstyle \left\langle v\right\rangle$, gives
\begin{equation}
\left\langle v\right\rangle =\frac{A_-+A_+}{\kappa} \quad.
\label{dryvel}
\end{equation}
Hence, in the dry friction limit, the particle moves in the direction
of the stronger of both strokes $A_\pm$. If $|A_+|>|A_-|$, the
particle moves into the positive direction while in the opposite case
it moves backwards.  Remarkably, $\left\langle v\right\rangle$ does
not explicitly depend on the (critical) friction force $\gamma$. No
matter how strong the stiff friction force $\gamma$ is, as long as
$\gamma<min(|A_-|,A_+|)$ is ensured, the average velocity
$\left\langle v\right\rangle$ does not depend on the friction.
Expressing the mean velocity (\ref{dryvel}) in terms of the three
effective parameters of the driving $p,A$ and $\tau$, we obtain
\begin{equation}
  \left\langle v\right\rangle= A\tau \frac{p(p-1)}{(p+1)^3}\quad ,
\label{dryvel1} 
\end{equation}
where we assumed, without loss of generality,
$\textstyle |A_-|>|A_+|$. Remember that we have per definition
$\textstyle 0<p<1$. Thus, in this case the current is negative.  
For a symmetric driving ($p=1$) the velocity
vanishes exactly. However, the velocity vanishes also if $\textstyle p \to
0$. That is why too much asymmetry is not a successful way to obtain
the optimum speed in the dry
friction limit. Therefore we
expect a maximum of the mean velocity within the 
region $\textstyle p\in[0,1]$. This result is confirmed by
the numerical simulations shown in Fig.(1).
\begin{figure}
  \begin{center}
 \includegraphics[width=12cm,height=9cm]{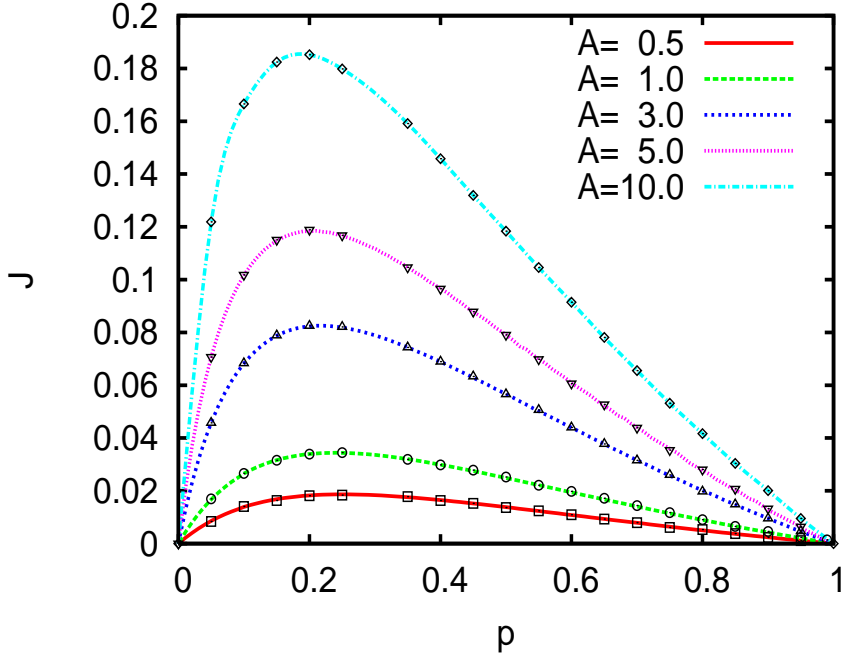}
 \caption{Case of dry friction: current versus asymmetry parameter $p$ for
   different stroke amplitudes $A$. The current shows a maximum for
   optimal asymmetry \cite{data}.}
\end{center}
\label{fig:agv}
\end{figure}

We now turn shortly to the case of Stokes friction ($n=1$ or 
$f(v)=-\gamma v$) and will take a look at the mean velocity.
For the stationary distribution $P_{st}^1$ we
compute
\begin{equation}
P_{st}^1(v)=N \ \lvert-\gamma v+A_+\rvert^{\frac{\kappa_+}{\gamma}-1} \ \lvert-\gamma v+A_-\rvert^{\frac{\kappa_-}{\gamma}-1}\quad.
\end{equation}
To determine the first moment $J=\left\langle v\right\rangle$, we get
\begin{equation}
\begin{split}
  \left\langle v\right\rangle=&\int_{\frac{A_-}{\gamma}}^{\frac{A_+}{\gamma}} v(-\gamma v+A_+)^{\frac{k_+}{\gamma}-1}(-\gamma v +A_-)^{\frac{\kappa_-}{\gamma}-1}dv\\
  &=\frac{c\Gamma[\frac{\kappa_-}{\gamma}]\Gamma[\frac{\kappa_+}{\gamma}]}{\Gamma[\frac{\kappa}{\gamma}+1]}\underbrace{\left(A_+\kappa_-+A_-\kappa_+\right)}_{=0}=0\quad.
\end{split}
\end{equation} 
Thus we have obtained, that in the linear case, there can never be a
current for vanishing mean forcing. The ratchet mechanism does not
work for linear friction.

In the limit of $n \to \infty$, there is no current as well. Since the relation (8) holds, there is no
friction forces in the interval $v\in[-1,1]$. 
Here $P_{st}^{\infty}(v)$ is constant. Hence, the first moment
vanishes and for the second moment we get $\left\langle v^2\right\rangle=1/3$.

\section{Stationary current: The general case}
In general cases ($n\in \mathbb{R}^+$), the analytical treatment is
difficult. Therefore we will turn to numerical measures and our
primary focus will lie on the mean velocity $\left\langle
  v\right\rangle$ which equals the stationary current. In addition, we
will look at the variance $var\left\lbrace v\right\rbrace $ and at the
inverse relative variance
\begin{equation}
 coh\lbrace v\rbrace= \frac{\left\langle v\right\rangle^2}{\left\langle v^2\right\rangle-\left\langle v\right\rangle^2}
\label{costud}
\end{equation}
which might serve as a measure of coherence of the transport
\cite{pikovsky}. High coherency means, that one has a low spread in
the transport process compared to the unidirectional drift.

\begin{figure}
  \begin{center}
    {\includegraphics[width=12cm,height=9cm]{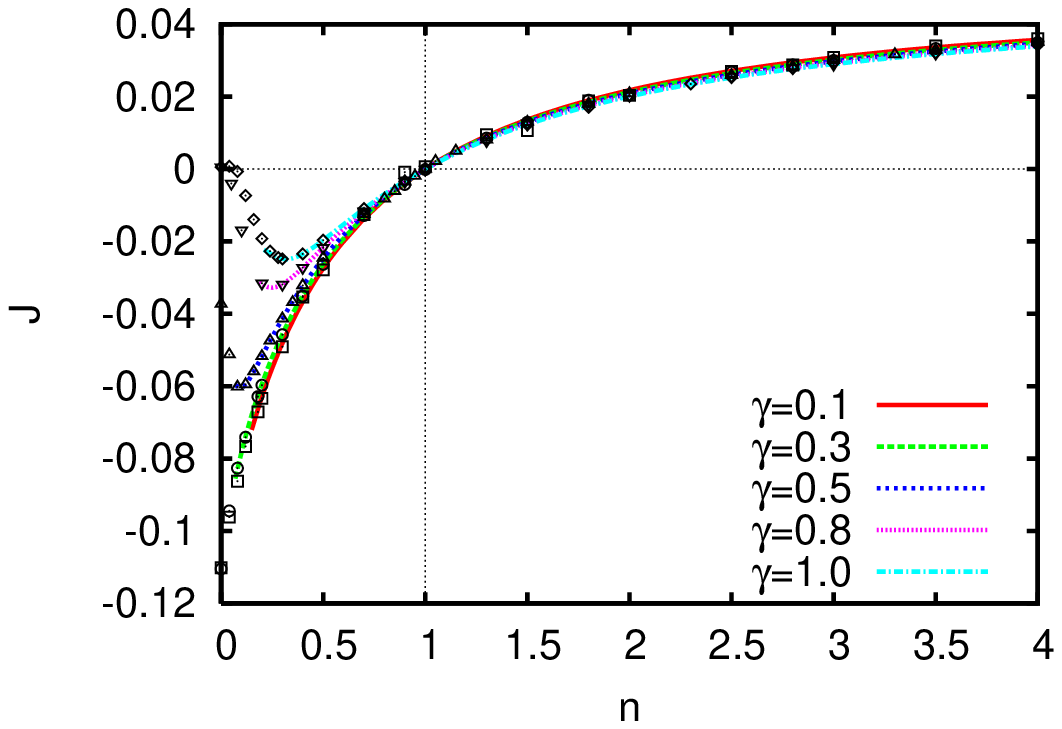}}
    {\includegraphics[width=12cm,height=9cm]{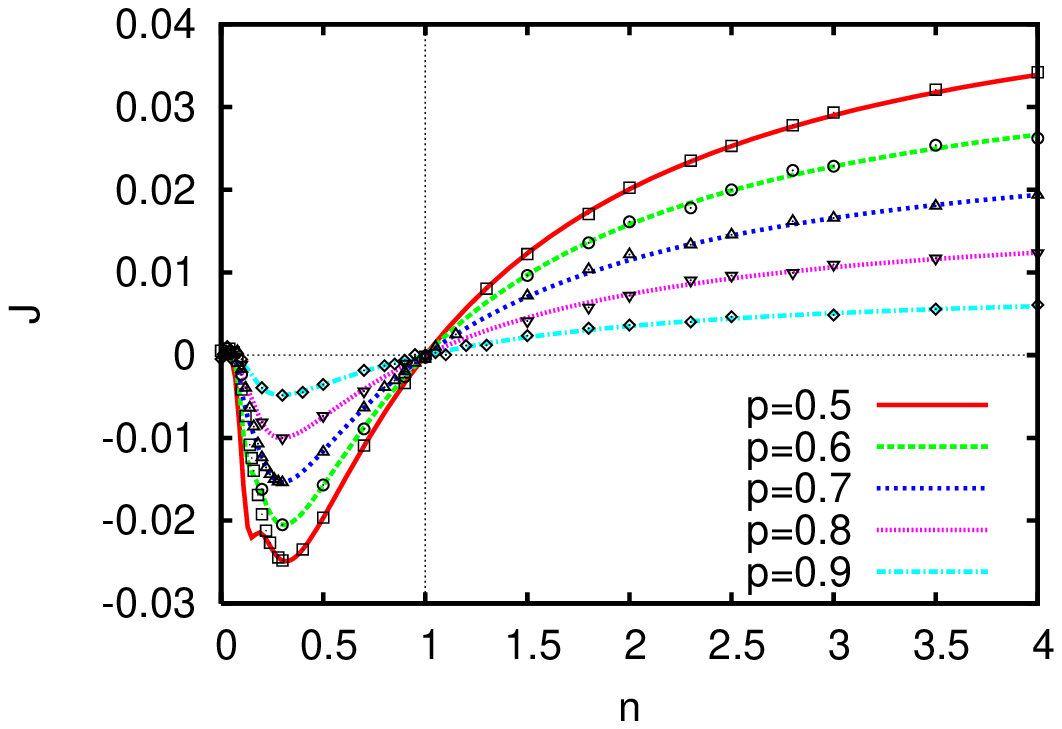}}
      \end{center}
\caption{Current $J$ versus friction exponent $n$ for different
  friction coefficients $\gamma$ (top) and driving asymmetries $p$ (bottom) \cite{data}.}
\label{non1}
\end{figure}

The impact of the nonlinearity parameter $n$ on the mean velocity
$\left\langle v\right\rangle $ is displayed in Fig.(\ref{non1}) (with $n\in[0,4]$)
for different realizations of the friction
coefficient $\gamma$ and the driving asymmetry $p$.
In all these cases we observe a single current reversal exactly at the
linear friction regime $n=1$. In other words, there is no transport
for linear friction. This is exactly the result we have found
analytically.  In particular, sublinear ($\textstyle n<1$) and
superlinear ($n>1$) friction results in opposing velocity directions.
While for sublinear friction, one observes a negative current, we get a
positive current for the superlinear friction regime.
Note that, without loss of generality, we have assumed $|A_-|>|A_+|$.
In the opposite case ($|A_+|>|A_-|$), one gets the same
picture with inverse sign. While the sublinear regime (dry friction regime)
results in the direction of the stronger amplitude, the
superlinear regime favors the opposite direction.
In the dry friction limit, we observe a different
behavior. If $\gamma<\max(|A_-|,|A_+|)$ holds, the
current becomes a non-zero value. Otherwise the current vanishes.
This is exactly what we predicted already in the analytical study of the
dry friction limit $n=0$. In the superlinear case the stationary
current does weakly depend on the friction. In the limit of large $n$,
the current will disappear.

\begin{figure}[tbh]
  \begin{center}
    {\includegraphics[width=12cm,height=9cm]{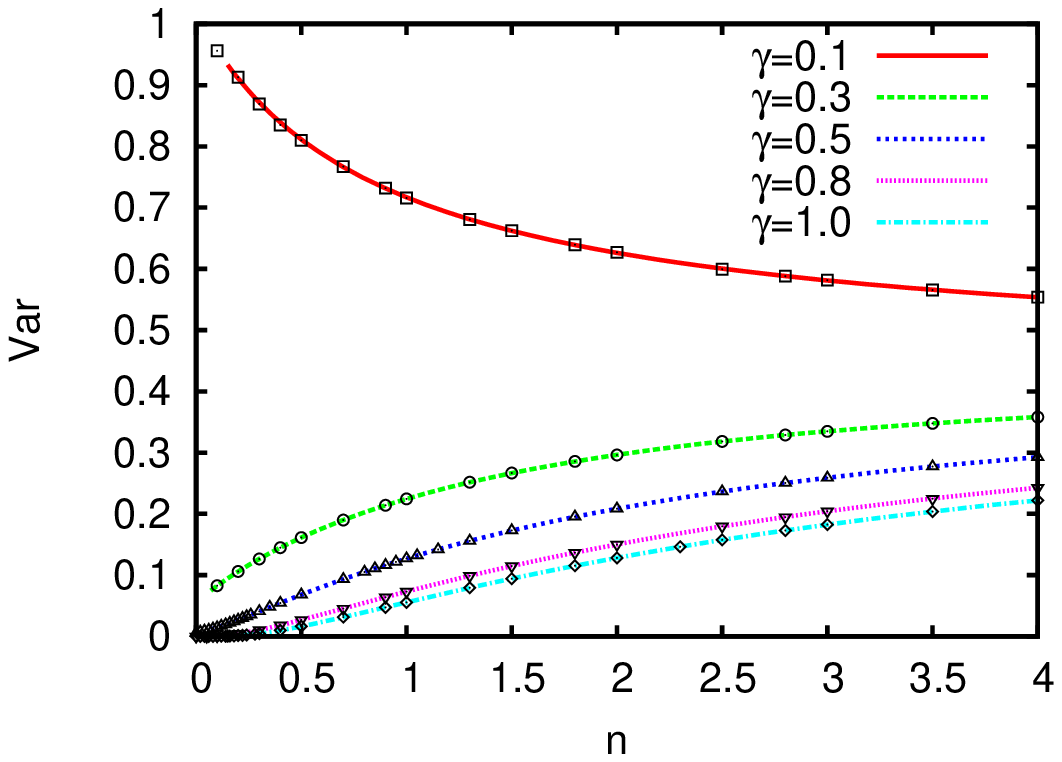}}
    {\includegraphics[width=12cm,height=9cm]{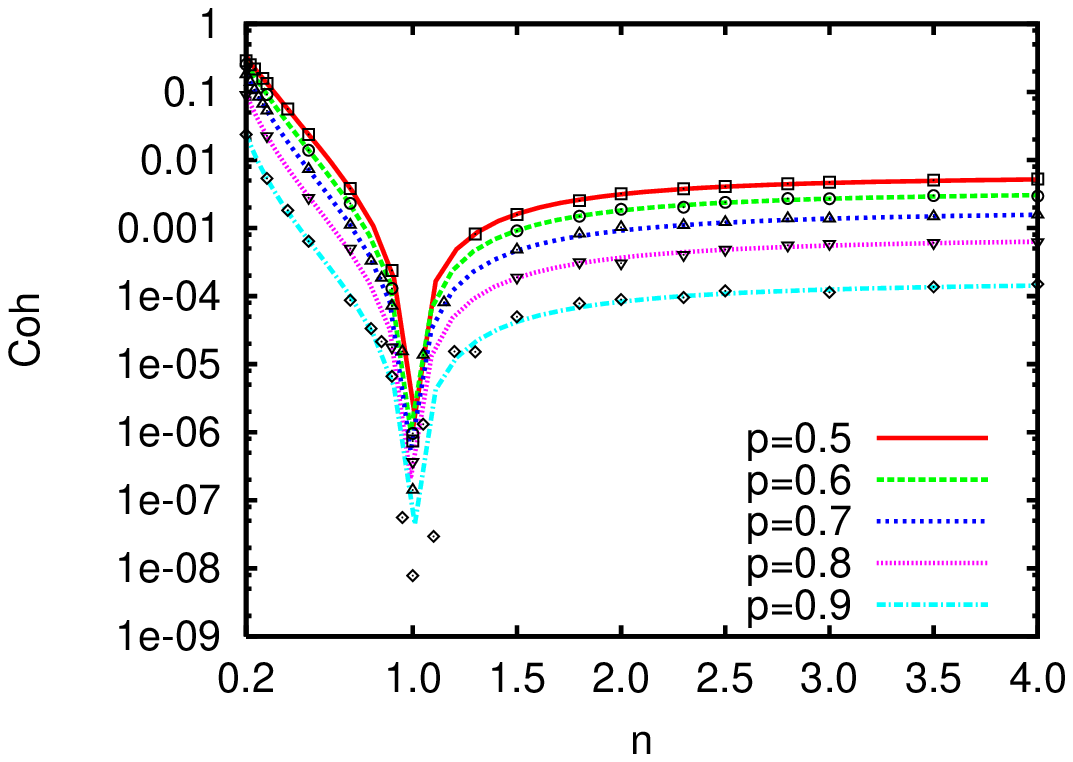}}
  \end{center}
  \caption{Variance of velocity for different friction
    coefficients $\gamma$ (top) and coherence (Eq.(\ref{costud})) for different driving
    asymmetries $p$ (bottom) versus friction exponent $n$ \cite{data}.}
\label{non2}
\end{figure}

The variance of the velocity is shown in Fig.(\ref{non2}).
When the critical force cannot be overcome by the
driving, the variance again vanishes in the dry friction limit. In
this situation the variance increases with growing friction exponent
$n$. In the case, where the critical force can be overcome by the
driving forces, we observe a finite variance for $n=0$, which decreases with increasing friction exponent.

The coherence is presented in Fig.(\ref{non2}). Since the current
$J$ vanishes for linear friction and at the same time the variance
keeps finite, we observe that the transport coherence tends to
vanish. We also see, that for superlinear friction ($n>1$), the
dependency of the coherence on the exponent $n$ is very weak.
$coh\lbrace v\rbrace$ almost does not change with varying friction exponent $n$.
Whereas we observe a rather strong
dependence on the exponent $n$ in the sublinear regime. In the limit $n\to 0$, the coherence
increases exponentially. Thus, the dry friction
regime shows much better coherence than the friction regime,
which occurs in fluids.

\begin{figure}[h]
  \begin{center}
    {\includegraphics[width=12cm,height=9cm]{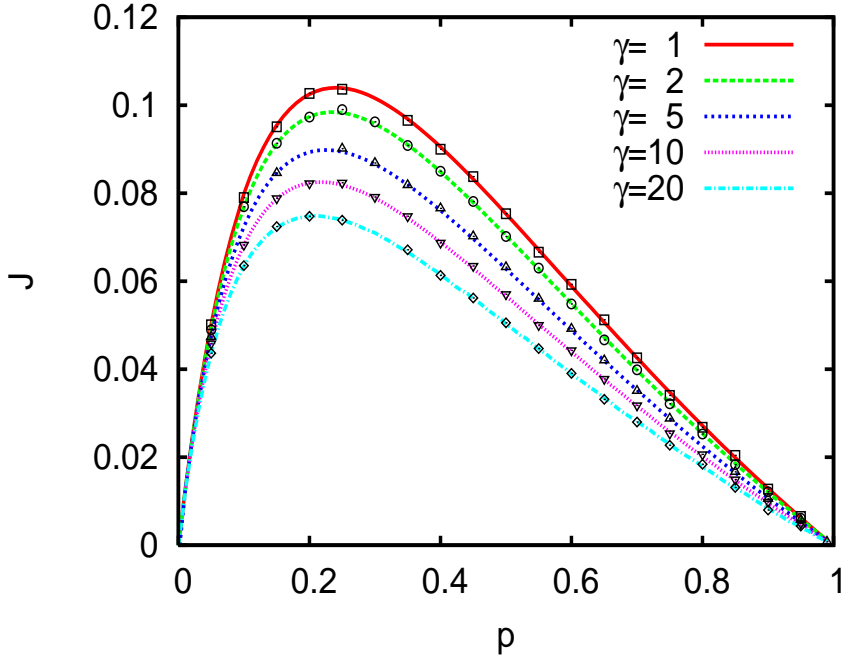}}
  \end{center}
  \caption{Current $J$ versus driving asymmetry $p$ for different
    friction coefficients $\gamma$ \cite{data}.}
  \label{sym4}
\end{figure}

Let us now investigate the asymmetry parameter $p$. In Fig.(\ref{sym4})
one observes in all cases a single maxima at symmetry
percentages below 0.5, varying slightly for different parameter
settings. In accordance with our previous considerations, the current
vanishes for a symmetric driving as well as in
the strongly asymmetric case. This is elucidated by
the specific values of driving amplitudes $A_{\pm}$ and
waiting times $\tau_{\pm}$ in this limit
\begin{align}
 \lim_{p\to0}A_-&=-A,\qquad\lim_{p\to0}A_+=0,
\label{symp1}\\
  \lim_{p\to0}\tau_-&=0,\qquad~~\;\;\lim_{p\to0}\tau_+=\tau \quad.
\end{align}
The average waiting time $\tau_-$ for the state $A_-$ vanishes.
Consequently, no negative force $A_-$ acts.
On the other hand, no positive force acts as well,
according to (25). Hence, no forces at all
are acting, if $p=0$.

\begin{figure}
  \begin{center}
    {\includegraphics[width=12cm,height=9cm]{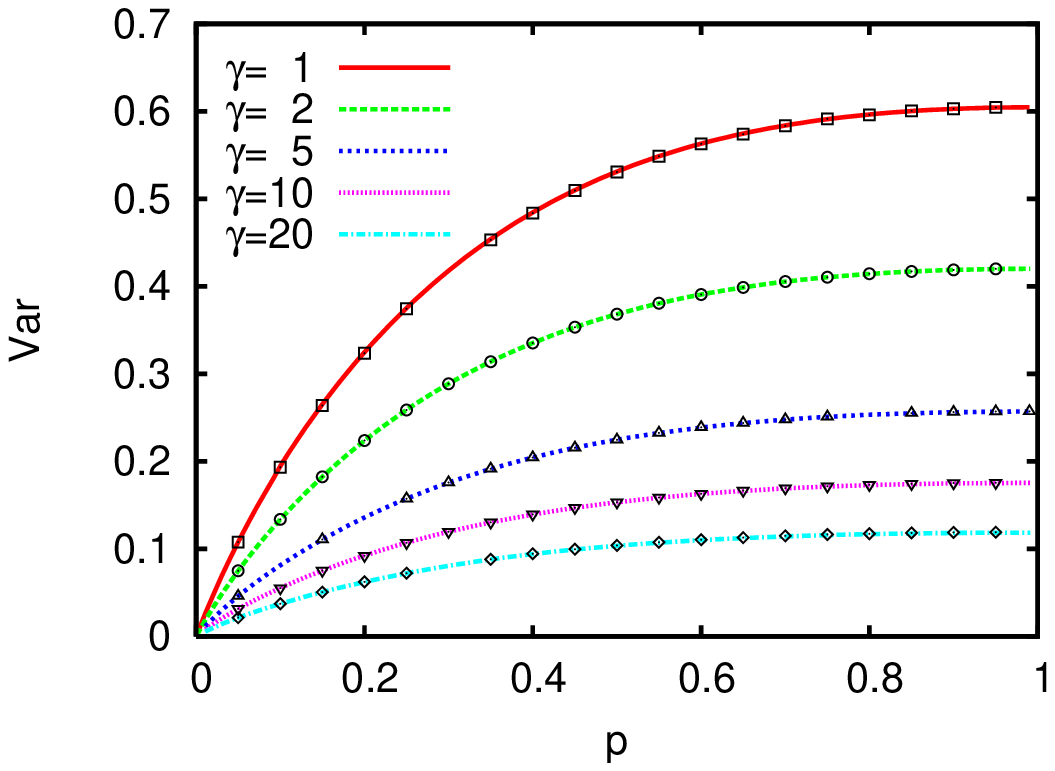}}
    {\includegraphics[width=12cm,height=9cm]{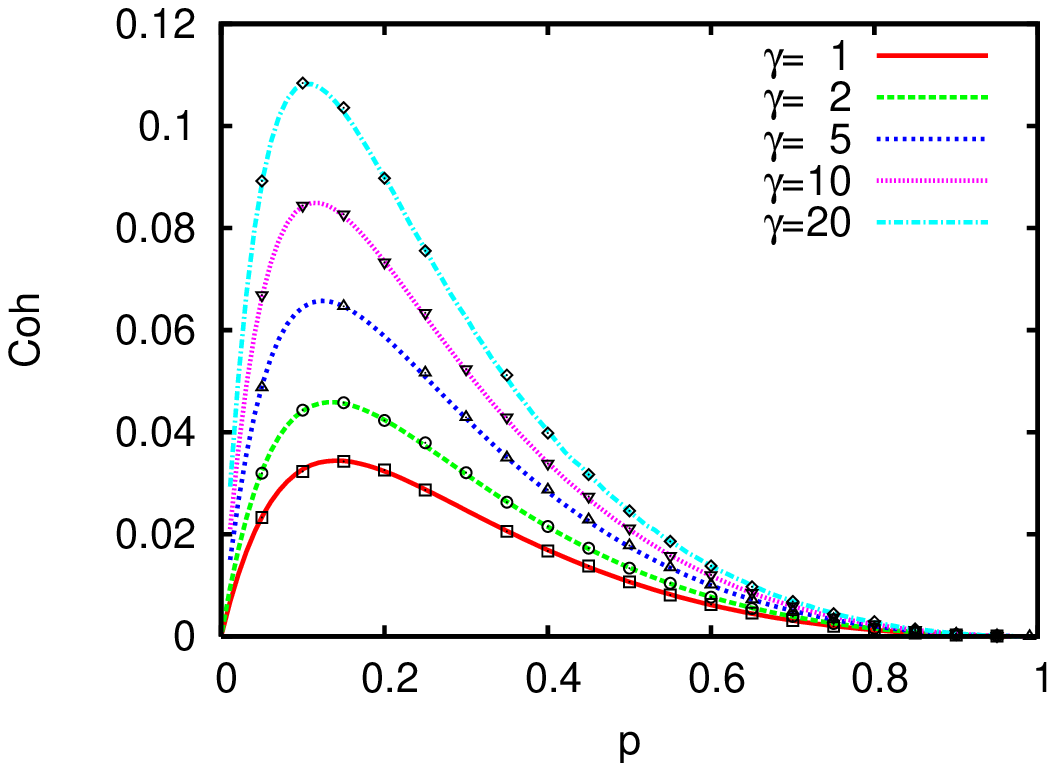}}
  \end{center}
  \caption{Variance (top) and coherence (Eq.(24);bottom) for different friction
    coefficients $\gamma$ versus driving asymmetry $p$ \cite{data}.}
\label{sym5}
\end{figure}

The same line of argument shows, that the variance also vanishes
in the asymmetric limit. Indeed, the results confirm
our ideas (Fig.(5)). Thus, starting at zero for $p=0$, the variance, in any
case, increases, until it reaches its maximum for $p=1$. This can be
understood in terms of the driving $\xi(t)$, which has already shown this
behavior (Eq. (6)). It seems reasonable that the variance of the state variable $v$ follows
qualitatively the variance of its driving.

The behavior of the coherence (Fig.(\ref{sym5})) shows similarities to
the behavior of the current $J$. One observes a single maximum, which
this time is shifted to higher asymmetries, i.e. to lower values of
$p$, compared to the location of the maxima of the current. For the regarded cases, the peaks
of the current are in regions where $p>0.2$
holds, while the peaks of the coherence are located below that
barrier. We observe a much better localization of the
maxima of the coherence, compared to those of the current.

\section{Diffusion in the two dimensional case}
Eventually, we present results for the two dimensional case. We
assume that the asymmetric drive $\xi(t)$ acts on the particle along a
certain variable direction in the two dimensional space. This
direction is given by the unit vector with Cartesian coordinates
\begin{equation}
  \label{eq:angle}
  \vec{e}_v(t) = {\cos\Theta(t) \choose \sin\Theta(t) }
\end{equation}
and $\Theta(t)$ is the present angle between the direction of the
drive and the $x$-coordinate. Along this axis, the particle changes
$v(t)$ with positive and negative values according to
Eq.(1) and
\begin{equation}
  \label{eq:veloc}
  \frac{{\rm d}}{{\rm d} t } \vec r(t)=\vec v(t)\,~~\mbox{while}~~  \vec v(t)=v(t)\cdot \vec e_v(t)\quad.
\end{equation}
Additionally, a Gaussian white noise source, labeled $\eta(t)$, rotates
the moving-axis of the particle. As equation for the angle $\Theta(t)$, we
formulate
\begin{equation}
  \label{eq:Theta}
  \frac{{\rm d}}{{\rm d} t } \Theta(t)=\frac{1}{v}\eta(t),  
\end{equation}
which assumes that the increment of the rotation angle scales with
$1/v$.  The noise $\eta(t)$ is defined by $\langle \eta(t) \rangle=0$
and $\langle \eta(t)\eta(t') \rangle= 2D_{\eta} \delta(t-t')$.

For the following simulations, we fixed the friction exponent $n=3$
with $|A_+|>|A_-|$, the intensity of the noise $D_{\eta}=0.1$ and all
other parameters as stated in \cite{data}.

\begin{figure}
  \begin{center}
  {\includegraphics[width=11.5cm]{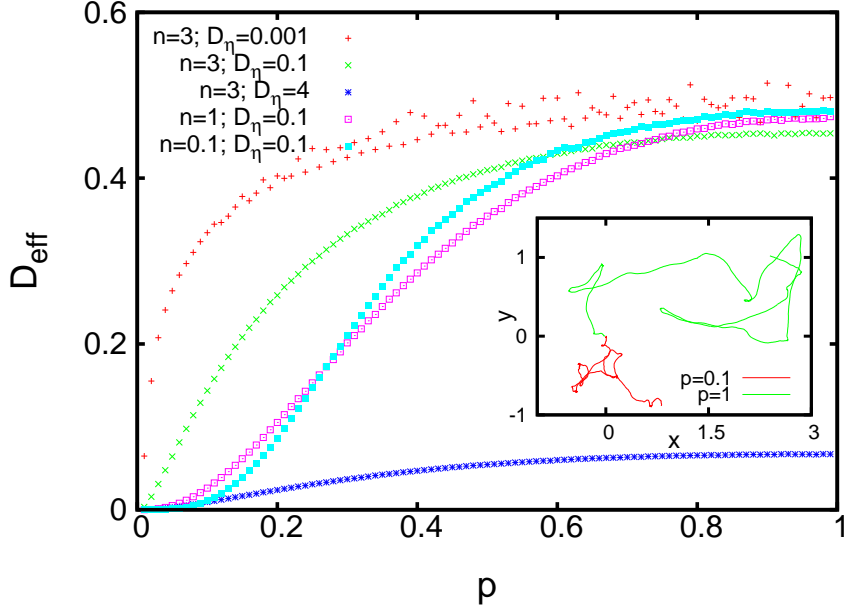}}
\end{center}
\caption{Diffusion coefficient of the two dimensional motion versus
driving asymmetry $p$ for several $D_{\eta}$ and
different friction exponents $n$; spatial trajectories for two different
driving asymmetries $p$ in the inset \cite{data}.}
\label{sym6}
\end{figure}


The motion of the particle shows a diffusive behavior after a crossover.
Due to the noisy rotations of the
asymmetric force, no preferred direction exists. Typical trajectories are presented in the
inset of Fig.(\ref{sym6}). Simulations show that the velocity distribution densities
have, for not too small $\tau$, a vanishing derivative at $v\to 0$ and possess two peaks
according to the asymmetric drive. This can be expected in regard to the one dimensional case
(\cite{sancho} and Eq.(9)). In the two dimensional velocity
space, two circles with high probabilities occur.

For large time scales, we assume
$\langle(\vec r(t)-\vec r(t_0))^2\rangle=4tD_{eff}$. The resulting behavior
of the diffusion coefficient (Fig.(\ref{sym6})) can be explained by
the trajectories in the inset. They illustrate the impact of the small
force $A_-$ which acts most of the time for $p=0.1$ and which leads to
a stronger influence of the angular noise, compared to the symmetric
case $p=1$.

As a possible explanation, we use the Taylor-Kubo-relation \cite{HaLsgIMFM07,LsgErdKom}
\begin{equation}
  \label{eq:kubo}
 \langle(\vec r(t)-\vec r(t_0))^2\rangle=2t\cdot \int_0^{\infty}C_{\vec v\vec v}(\tau)\mathrm d\tau\quad. 
\end{equation}
In simulations we have inspected that the velocity correlations
$C_{\vec v\vec v}(\tau)$ increases for stronger symmetries $p$ in the
driving due to an increased standard deviation of the velocities.
Therefore the diffusion coefficient $D_{eff}$ is maximal for symmetric noise.

The slowly growing diffusion for small friction exponents and small
driving symmetries $p$ (Fig.(\ref{sym6})), can be explained by the
nearly vanishing velocity $v_-$, which mostly acts in this limit and
depends on $A_-$.  We also observe that larger noises $D_{\eta}$ lead
to smaller diffusion coefficients.  This behavior has already been
studied in similar systems \cite{HaLsgIMFM07,MiMe} and is caused
by a straight movement for small angular noise.  The
analytical treatment of the diffusion coefficient, for the one
dimensional DMN process, has been recently studied for a symmetric
driving \cite{Lin09} and corresponds to our results for $p=1$ in the
limit of small noises $D_{\eta}$.

\section{Conclusion and outlook}
We have shown that directed motion in nonlinear media is possible, if
particles are driven by an asymmetric force. The force can have
external or internal origin. Small additive noise would only partially
destroy the obtained effect as was seen in calculation of the
stationary distribution densities \cite{dybiec}. In the one
dimensional case, it results in a smoothing of the densities but the
averaged motion will survive. Diffusion will be obtained with
properties similar to the recently studied case with symmetric drive
\cite{Lin09}.

While the analytically and numerically studied one dimensional case is
well understood, the behavior of particles in two dimensions remains
still a challenge. Also further energetic consideration will be
helpful in order to estimate the efficiency of the proposed ratchet.
Future fields of interest will include also alternative two
dimensional models for DMN processes and the influence of coupling
forces to external obstacles or mutually to similar objects. We also
see a great challenge in the consideration of different friction terms
or periodically changing friction forces in order to simulate rhythmic
movements in biological systems.

\section{Acknowledgments}
This paper was supported by DFG-Sfb 555 ``Nonlinear complex
processes'' and by VW-Foundation ``New conceptional approaches to
modeling and simulating complex systems''. We acknowledge help by
Jessica Strefler who has been previously involved in this study.


\begin{thebibliography}{0}
\bibitem{data}The default values are $\textstyle \tau=1.5, A=3, p=0.5,
 n=3, \gamma=1$, meaning that, when a parameter is not specified in
  the plot, then its value is the default value.
\end{thebibliography}
\end{document}